\documentclass{article}
\usepackage{waspaa17,amsmath,graphicx,url,times,multirow}

\usepackage{color}
\usepackage{cite}

\title{Multi-scale Multi-band DenseNets for Audio Source Separation}

\name{Naoya Takahashi, Yuki Mitsufuji}

\address{Sony Corporation, Minato-ku, Tokyo, Japan}

\begin{document}

\ninept
\maketitle

\begin{sloppy}

\begin{abstract}
 
This paper deals with the problem of audio source separation. To handle the complex and ill-posed nature of the problems of audio source separation, the current state-of-the-art approaches employ deep neural networks to obtain instrumental spectra from a mixture. In this study, we propose a novel network architecture that extends the recently developed densely connected convolutional network (DenseNet), which has shown excellent results on image classification tasks. To deal with the specific problem of audio source separation, an up-sampling layer, block skip connection and band-dedicated dense blocks are incorporated on top of DenseNet. The proposed approach takes advantage of long contextual information and outperforms state-of-the-art results on SiSEC 2016 competition by a large margin in terms of signal-to-distortion ratio. Moreover, the proposed architecture requires significantly fewer parameters and considerably less training time compared with other methods.
\end{abstract}


\section{Introduction}
\label{sec:intro}

Audio source separation has attracted considerable attention in the last decade. Various approaches have been introduced so far such as local Gaussian modeling \cite{DuongVG10, FitzgeraldLB16}, non-negative factorization \cite{LiutkusFB15, Roux15, MitsufujiKS16}, kernel additive modeling \cite{LiutkusFRPD14} and combinations of those approaches \cite{OzerovF10, LiutkusFR15, FitzgeraldLB162}. 
Recently, deep neural networks (DNNs) based source separation methods has shown significant improvement in separation performance over earlier methods.
In \cite{Nugraha15,Uhlich15}, a standard feed-forward fully connected network (FNN) was used to obtain the source spectra. As an input for the FNN, multiple frames (typically up to about 20 frames) were concatenated to take advantage of temporal contexts. 
To model longer contexts, long short term memory (LSTM) was used in \cite{Uhlich17}. Despite its good performance, the LSTM usually requires a relatively long training time, making it difficult to re-train the network to adapt to different domains or to explore the best architecture.

Another well-known architecture, Convolutional Neural Network (CNN)~\cite{LeCun1998} has been very successful in image domain and also widely used in a variety of audio and video tasks\cite{Sercu2015, Takahashi2016, Korzeniowski16, Takahashi2017}. As the convolution layers are stacked, a receptive field of the deeper layer covers a larger area of the input field, enabling the deep CNN architecture to take long contexts, as LSTM does. However, considerable depth is required to cover long contexts, making the network training difficult and leading to performance degradation \cite{He2016}.
Recent works, such as ResNets \cite{He2016} and Highway Networks \cite{Srivastava15}, address this problem by bypassing signals from one layer to the next via identity connections; this enable to successfully train the networks with more than 100 layers. Most recently, a novel CNN architecture called densely connected convolutional networks (DenseNet) has shown excellent performance on image recognition task \cite{Huang2016}. The idea of DenseNet is to use concatenation of output feature maps of preceding layers as the input to succeeding layers. Unlike ResNet, this iterative connection enables the network to learn explicit cross-layer interactions and reuses features computed in preceding layers, which yields efficient use of parameters. This property suits the audio source separation problem very well because the goal of audio source separation is to estimate the instrumental spectrograms buried in interference sounds and the estimated source spectrograms could be brushed up more easily by referring the mixture or previous layer outputs. 
However, DenseNet is inherently memory demanding because the number of inter-layer connections grows quadratically with depth. Even for image recognition tasks involving relatively low resolution images (for instance 32 $\times~$32 input and 10 or 100 output, as in CIFAR \cite{Krizhevsky09} and SVHN \cite{Netzer11}), the authors used pooling layers to overcome the explosion of the number of feature maps. In audio source separation, both the input and output dimension would be far larger (e.g. 1024 frequency bins $\times$ 128 frames) in order to utilize sufficiently long contexts with high frequency resolution.

To address this problem, we propose a fully convolutional multi-scale DenseNet equipping {\it dense blocks} with multiple resolutions. The input of lower resolution dense blocks is created by iteratively down-sampling the outputs from the preceding dense blocks. The low resolution dense blocks are then up-sampled to recover a higher resolution, and the result are fed to higher resolution dense blocks together with the output from the preceding dense blocks having same resolution, as shown in Fig. \ref{fig:mdense}. The lower resolution blocks capture the entire context while the higher resolution blocks recover details of the time-frequency structure in the spectrogram. This architecture enables the network to model both long contexts and fine-grained structures efficiently within a practical model size while maintaining the advantages of DenseNet.
In order to increase the modeling capability, we further introduced dense blocks dedicated to particular frequency bands as well as to the entire frequency spectrum. 
Although convolution along the frequency axis is shown to be effective in the audio domain including speech \cite{Sercu2015} and non-speech \cite{Takahashi2016}, local patterns in the spectrogram are often different in different frequency bands: the lower frequency band is more likely to contain high energies, tonalities and long sustained sounds, whereas the higher frequency band tends to have low energies, noise and rapidly decaying sounds. Most kernels in a convolution layer focus on the higher energy band and neglect the lower energy band, which they consequently fail to recover. Therefore, we propose dense blocks dedicated to each band. In combination with a global dense block, the network is thus able to model efficiently both local and global structures.

The contributions of this paper are as follows:
\begin{enumerate}
\item We propose multi-scale fully convolutional networks for audio source separation by extending DenseNet to cover long contexts while enabling the network to model large input and output dimensions.

\item We further propose to model each frequency band separately, enabling the kernels to focus on particular distribution, which differ for each frequency band.

\item The proposed method largely outperforms the state of the art that achieved the best score in the Signal Separation Evaluation Campaign (SiSEC) 2016 competition \cite{Liutkus17}. Moreover, it considerably reduces the training time and the number of parameters in comparison with recently proposed DNN based methods.
\end{enumerate}


\section{Multi-scale multi-band DenseNet}
In this section, we first summarize the DenseNet architecture. Then, we extend DenseNet by introducing up-scaling blocks and inter block skip connections to deal with the high dimensional inputs and outputs that are inherent to utilize long context with high resolution audio. Next, we introduce a multi-band DenseNet architecture that improves modeling efficiency and capability. Finally, the complete architectures are outlined.

\subsection{DenseNet}
\label{sec:DenseNet}

\begin{figure}[t]
\centering
\includegraphics[width=77mm]{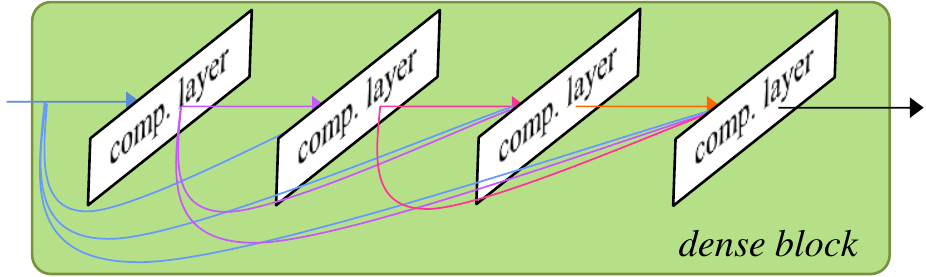}
\caption{dense block architecture. The input of a composite layer is the concatenation of outputs of all preceding layers.}
\label{fig:db}
\end{figure}

In a standard feed forward network, the output of the $l$th layer is computed as $x_l = H_l(x_{l-1}),$
where the network input is denoted as $x_0$ and $H_l(\cdot)$ is a non-linear transformation which can be a composite function of operations such as Batch Normalization (BN) \cite{Ioffe15}, rectified linear units (ReLU) \cite{Glorot11}, pooling, or convolution. In order to mitigate difficulties of training very deep models, ResNet \cite{He2016} employs a skip connection which adds an identity mapping of the input to the non-linear transformation:
\begin{equation}
x_l = H_l(x_{l-1}) + x_{l-1}.
\label{eqResNet}
\end{equation}
The skip connection allows the network to propagate the gradient directly to the preceding layers, making the training of deep architectures easier. 
DenseNet \cite{Huang2016} further improves the information flow between layers by replacing the simple addition of the output of a single preceding layer with a concatenation of all preceding layers:
\begin{equation}
x_l = H_l([x_{l-1}, x_{l-2},\ldots,x_0]),
\label{eqDenseNet}
\end{equation}
where $[\ldots]$ denotes the concatenation operation. Such dense connectivity  enables all layers not only to receive the gradient directly but also to reuse features computed in preceding layers. This avoids the re-calculation of similar features in different layers, making the network highly parameter efficient. Fig. \ref{fig:db} illustrate the dense block.
In DenseNet, $H_l$ comprises of BN, followed by ReLU and convolution with $k$ feature maps. In the reminder of this paper, $k$ is referred to as {\it growth rate} since the number of input feature maps grows linearly with depth in proportion to $k$ (e.g. the input of $l$th layer have $l\times k$ feature maps).

For image recognition tasks, a pooling layer, which aggregates local activation and maps to the lower dimension, is essential to capture the global information efficiently. A {\it down-sampling layer} defined as a 1 $\times$~1 convolution followed by a 2 $\times$~2 average pooling layer is introduced to facilitate pooling. By alternately connecting dense blocks and down-sampling layers, the feature map dimension is successively reduced and finally fed to a softmax classification layer after global pooling layer. In the next section, We discuss how to apply these ideas to audio source separation.

\subsection{Multi-Scale DenseNet with block skip connection and transposed convolution}
\label{sec:MDenseNet}
\begin{figure}[t]
\centering
\includegraphics[width=\linewidth]{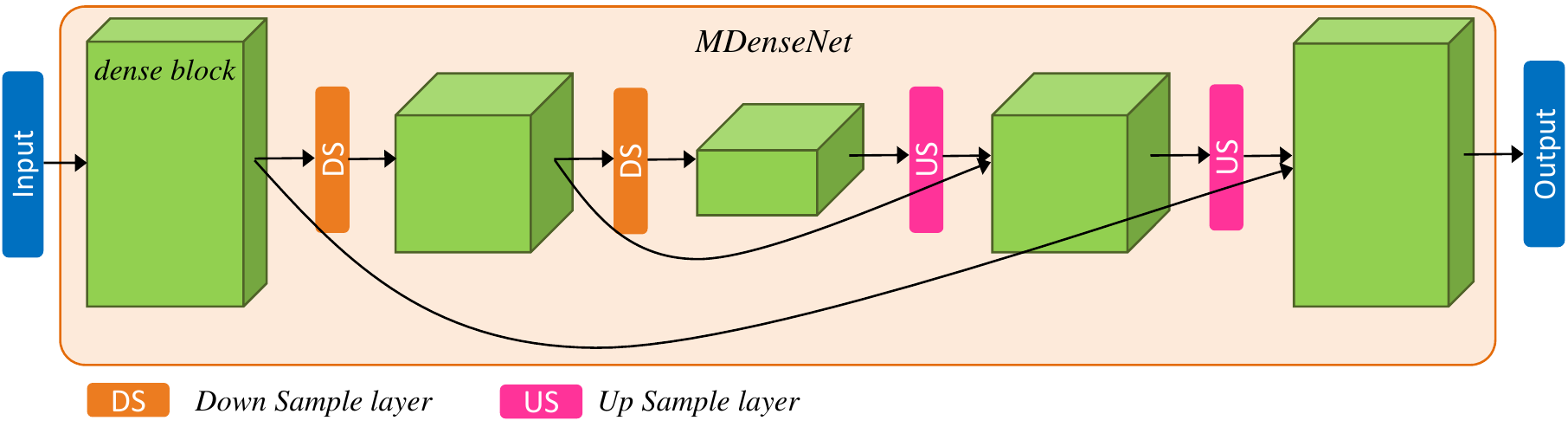}
\caption{MDenseNet architecture. Multi-scale dense blocks are connected though down- or up-sampling layer or through block skip connections. The figure shows the case $s=3$.}
\label{fig:mdense}
\end{figure}

Dense blocks and down-sampling layers comprise the down-sampling path of the proposed multi-scale DenseNet. Down-sampled feature maps enable the dense block network to model longer contexts and wider frequency range dependency while alleviating computational expense. 
In order to recover the original resolution from lower resolution feature maps, we introduce an up-sampling layer defined as a transposed convolution whose filter size is same as the pooling size. We again alternate up-sampling layers and dense blocks to successively recover the higher resolution feature maps. In order to allow forward and backward signal flow without passing though lower resolution blocks, we also introduce {\it inter-block skip connection} which directly connect two dense blocks of the same scale. With this connection, dense blocks in the down-sampling path are enabled to receive supervision and send the extracted features without compressing and decompressing them. The idea of the entire architecture is depicted in Fig. \ref{fig:mdense} in case that the number of different scales $s$ is 3 which can be tuned depends on a data complexity and resource availabilities. Hereafter, we refer to this architecture as {\it MDenseNet}. Note that the proposed architecture is fully convolutional and thus can be applied to arbitrary input length.

\subsection{Multi-band MDenseNet}
\label{sec:MMDenseNet}
\begin{figure}[t]
\centering
\includegraphics[width=\linewidth]{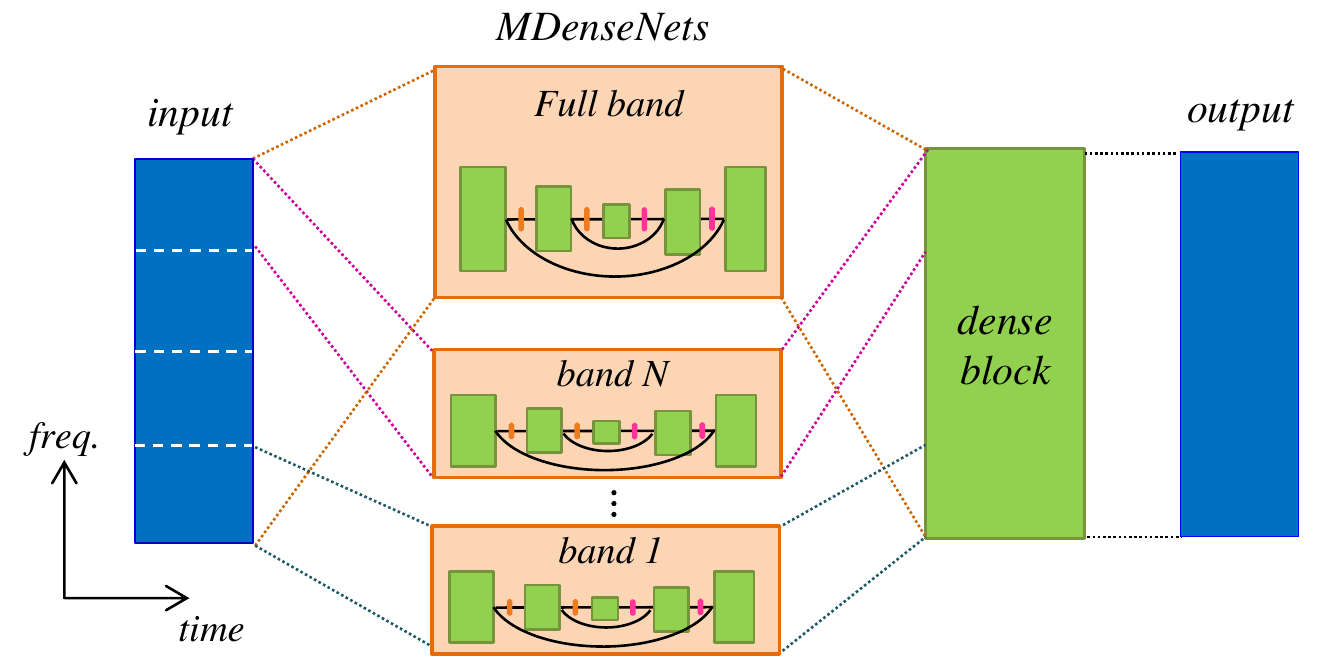}
\caption{MMDenseNet architecture. Outputs of MDenseNets dedicated for each frequency band including full band are concatenated and the final dense block integrate features from these bands to create final output.}
\label{fig:mmdense}
\end{figure}

In the architecture discussed in Sec.~\ref{sec:MDenseNet}, the kernels of the convolution layer are shared across the entire input field. This is reasonable if the local input patterns appear in any position in the input, as is the case for objects in natural photos. In audio, however, different patterns occur in different frequency bands, though a certain amount of translation of patterns exists, depending on the relatively small pitch shift. Therefore, limiting the band that share the kernels is more suitable for efficiently capturing local patterns. Indeed, limited kernel sharing has been shown to be effective in speech recognition \cite{Abdel-Hamid2013}. We split the input into multiple bands and apply multi-scale DenseNet to each band. However, simply splitting frequency band and modeling each band individually may hinder the ability to model the entire structure of spectrogram. Hence, we build in parallel an MDenseNet for the full band input and concatenate its output with outputs from multiple sub-band MDenseNets, as shown in Fig. \ref{fig:mmdense}. Note that in this architecture, since fine structure can be captured by band limited MDenseNets, the full band MDenseNet can focus on modeling rough global structure, thus simpler and less expensive model can be used. We refer to the architecture as MMDenseNet.

\subsection{Architecture details}
\label{sec:detailArch}
Details of the proposed network architectures for audio source separation are described in Table \ref{tab:densearch}. One advantage of MMDenseNet is that we can design suitable architectures for each band individually and assign computational resources according to the importance of each band which may differ depending on the target source or application. In this work, we split the frequency into two bands in the middle and design a relatively larger model for the lower frequency band.

 \begin{table}[t]
    \caption{\label{tab:densearch} {\it The proposed architectures. All dense blocks are equipped with 3$\times$3 kernels with $L$ layers and $k$ growth rate. The pooling size and transposed convolution kernel size are 2$\times$2. }}
    \vspace{2mm}
    \centerline{
      \footnotesize
      \tabcolsep=3px
      \begin{tabular}{ c | c | c | c | c | c } 
        \hline
        \multirow{2}{*}{Layer} & \multirow{2}{*}{scale} & \multicolumn{3}{c|}{MMDenseNet} & \multirow{2}{*}{MDenseNet}\\
        \cline{3-5}
        & & low & high & full &  \\
        \hline
        band split & \multirow{3}{*}{1} & first half & last half &  - & -  \\
        conv (t$\times$f,ch) &  & 3$\times$4, 32 & 3$\times$3, 32  & 3$\times$4, 32 & 3$\times$4, 32 \\
        dense 1 (k,L) & & 14, 4 & 10, 3 & 6, 2 & 12, 4 \\
        \hline                 
        down sample  & \multirow{2}{*}{$\frac{1}{2}$} & pool & pool  & pool & pool \\
        dense 2 (k,L) & & 16, 4 & 10, 3 & 6, 2 & 12, 4 \\
        \hline
        down sample  & \multirow{2}{*}{$\frac{1}{4}$} & pool & pool & pool & pool\\
        dense 3 (k,L) & & 16, 4 & 10, 3 & 6, 2 & 12, 4 \\
        \hline
        down sample  & \multirow{2}{*}{$\frac{1}{8}$} & pool & pool & pool & pool \\
        dense 4 (k,L) & & 16, 4 & 10, 3 & 6, 4 & 12, 4 \\
        \hline
        up sample  & \multirow{3}{*}{$\frac{1}{4}$} & t.conv  & t.conv  & t.conv & t.conv \\
        concat. & & low dense 3 & high dense 3 & full dense 3 & dense 3 \\
        dense 5 (k,L) & & 16, 4 & 10, 3 & 6, 2 & 12, 4 \\
		\hline
        up sample  & \multirow{3}{*}{$\frac{1}{2}$} & t.conv & t.conv & t.conv & t.conv \\
        concat. & & low dense 2 & high dense 2 & full dense 2 & dense 2 \\
        dense 6 (k,L) & & 16, 4 & 10, 3 & 6, 2 & 12, 4 \\
		\hline
        up sample  & \multirow{3}{*}{1} & t.conv & t.conv & t.conv & t.conv \\
        concat. & & low dense 1 & high dense 1 & full dense 1 & dense 1 \\
        dense 7 (k,L) & & 16, 4 & 10, 3 & 6, 2 & 12, 4 \\        
        \hline
        concat. (axis) & \multirow{4}{*}{1} & \multicolumn{2}{c|}{freq} & - & \multirow{2}{*}{-} \\
        \cline{3-5}
        concat. (axis) &  & \multicolumn{3}{c|}{channel} &  \\
        \cline{3-6}
        dense 8 (k,L) & & \multicolumn{3}{c|}{4, 2} & 4, 2 \\   
        \cline{3-6}
        conv(t$\times$f,ch) &   &   \multicolumn{3}{c|}{1$\times$2, 2} & 1$\times$2, 2 \\
        \hline
      \end{tabular}
    }
\end{table}

\section{Experiments}
\subsection{Setup}
\label{sec:setup}
We evaluated our proposed method on {\it DSD100} dataset which is build for SiSEC \cite{Liutkus17}. The dataset consists of {\it Dev} and {\it Test} sets with 50 songs each, recorded in stereo format at 44.1kHz sampling frequency. The average duration of songs is about 4 minutes. For each song, the mixture and its four sources, {\it bass, drums, other} and {\it vocals}, are available. The task is to separate songs into the four source instruments, or simply into  the {\it vocals} and {\it accompaniment} track. 
We used a spectrogram (sequence of short-time Fourier transform (STFT) magnitudes obtained by using a frame size of 2048 samples with 50\% overlap) of the mixture $X(t,f)$ as the input and trained a network to estimate target spectrogram $S_i(t,f)$ by minimizing the square error between the network output $\hat{S}_i(t,f)$ and $S_i(t,f)$, where $f$ is the frequency bin index, $t$ is the frame index and $i \in \mathcal{I}=\{bass, drums, others, vocals\}$ is the index of instruments.
The training was conducted with RMSprop \cite{Tieleman2012}, with an initial learning rate of 0.001 and reduced to 0.0001 after the performance saturated. Networks were trained individually for each instrument using data augmentation and the estimates $\hat{S}_i(t,f)$ were further enhanced by applying multi-channel Wiener filter (MWF), as in \cite{Uhlich17}. 

\subsection{State of the art comparison}
\begin{figure*}[t]
\centering
\includegraphics[width=175mm]{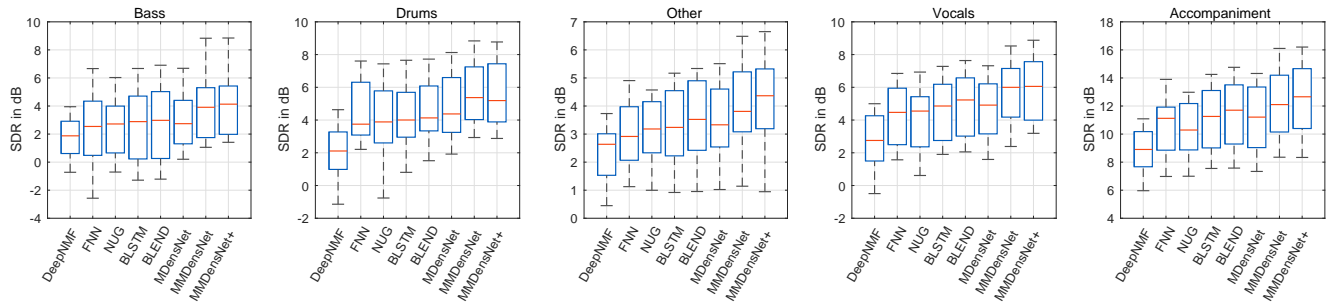}
\caption{{\it SDR comparison. Red line indicates the median and blue box indicates the 50\% percentile.}}
\label{fig:boxplot}
\end{figure*}

\begin{table}[t]
\caption{\label{tab:ex1} {\it Comparison of SDR.}}
\vspace{2mm}
\centering{
\begin{tabular}{ c | c c c c c} 
\hline
\multicolumn{1}{c|}{} & \multicolumn{5}{c}{SDR in dB}\\
Method      &	Bass	&	Drums	& Other & Vocals & Acco.	\\
\hline\hline
DeepNMF \cite{Roux15}	&	1.88	& 2.11 & 2.64 & 2.75 &  8.90 \\
NUG \cite{Nugraha15}\	&	2.72	& 3.89 & 3.18 & 4.55 & 10.29 \\
FNN \cite{Uhlich15}\	&	2.54	& 3.75 & 2.92 & 4.47 & 11.12 \\
BLSTM \cite{Uhlich17}\	&	2.89	& 4.00 & 3.24 & 4.86 & 11.26 \\
BLEND \cite{Uhlich17}\	&	2.98	& 4.13 & 3.52 & 5.23 & 11.70 \\
\hline
MDenseNet 	&	2.74	& 4.37 & 3.33 & 4.91 & 11.21 \\
MMDenseNet 	&	3.91	& {\bf 5.37} & 3.81 & 6.00 & 12.10 \\
{\bf MMDenseNet+} 	&	{\bf 4.13}	& 5.19 & {\bf 4.37} & {\bf 6.06} & {\bf 12.66}\\

\hline
\end{tabular}
}
\end{table}  

We compared our method with other state-of-the-art approaches:
\begin{itemize}
\item DeepNMF \cite{Roux15}: Non-negative deep network architecture which results from unfolding NMF iterations and untying their parameters.
\item NUG \cite{Nugraha15}: This approach estimates source spectra using DNN, and iteratively updates the spatial and spectral estimates using expectation-maximization. This approach was referred as NUG1 in \cite{Nugraha15}.
\item FNN \cite{Uhlich15}: The source spectra was estimated by feed forward fully connected DNN trained with an additional dataset (MedleyDB \cite{Bittner14}). Final outputs were obtained by applying single-channel Wiener filter to each channel individually. 
\item BLSTM \cite{Uhlich17}: Three layer bidirectional long short time memory (BLSTM) was used to estimate source spectrogram. This system marked second best score in SiSEC 2016 competition \cite{Liutkus17} and can be considered as a good baseline since it also uses MWF, thus the performance difference between these system highlight the effect of our proposed network architectures.
\item BLEND \cite{Uhlich17}: This approach linearly blend the estimates of FNN and BLSTM before applying MWF. The best score on SiSEC 2016 competition was obtained with this approach.
\end{itemize}

Table \ref{tab:ex1} and Fig. \ref{fig:boxplot} show  the signal to distortion ratio (SDR) computed using the BSS Eval toolbox \cite{Vincent06}. Among the state-of-the-art baselines, BLEND showed the best performance, which was a fusion of BLSTM and FNN.
MDenseNet performed as good as BLSTM, which also utilized MWF. This suggests that the multi-scale architecture successfully learned to utilize long term contexts using the stack of convolution layers instead of the recurrent architecture. This claim will be further investigated in the next subsection. MMDenseNet significantly improved performance and largely outperformed all baselines, showing the effectiveness of the multi-band architecture. We also trained MMDenseNet with the additional dataset, MedleyDB as FNN approach, and denoted it as MMDenseNet+. It further improved performances for all instruments except {\it drums} and showed the best overall result. Notably, we obtained 0.97dB improvement on average over the best results of the SiSEC 2016.         
        
\subsection{Architecture validation}
\begin{figure}[t]
\centering
\includegraphics[width=\linewidth]{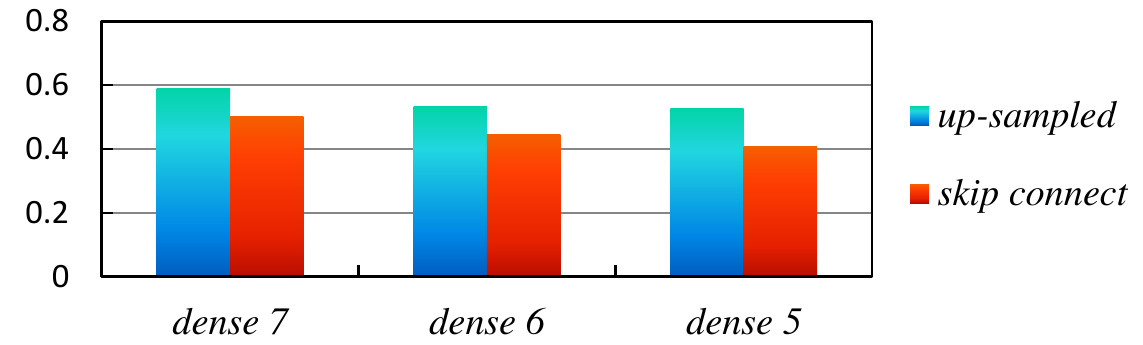}
\caption{The average norm of kernels for the skip connection path and the up-sampled path.}
\label{fig:l2norm}
\end{figure}

The proposed multi-sale dense block enables the network to model the signal on different scales, i.e. the global context in the down-scaled blocks and local fine-grained structure in the high resolution blocks. To validate if dense blocks at each scale actually contribute to recovering the target spectrogram, we computed the map-wise l2-norm of filter weights of dense blocks in up-sampling path (dense 5, 6 and 7 in Table \ref{tab:densearch}). The input of dense blocks in the up-sampling path is the concatenation of the output of the preceding up-sampling layer from down scaled block, and the skip connection from the dense block in down-sampling path, as in Fig. \ref{fig:mdense}. By comparing the averaged l2-norm of the filter weights corresponding to the up-sampling path and the skip connection path, we can conjecture the contribution of dense blocks in different scale. Fig.\ref{fig:l2norm} shows that the l2-norms of these two path are roughly the same, indicating that every dense block at different scale indeed contributes reasonably. This validate the advantage of the multi-scale DenseNet structure.

\subsection{Model efficiency}
\label{sec:eff}
The proposed architecture encourages feature reuse within and between dense blocks, leading to a compact and efficient model. To verify this, the number of parameters and the model training times are compared in Table \ref{tab:eff}. The number of parameters of the proposed architectures are significantly less than the baseline methods. MDenseNet achieved comparable performance to the sate-of-the-art BLEND approach with only 1.5\% of the parameters, and MMDenseNet largely outperformed BLEND with only 3.6\% of the parameters. This demonstrates the compactness and efficiency of the model, which is preferable for deployment. The training time is also significantly less than for the BLSTM and BLEND methods, making it easier to tune the hyper-parameters.

\begin{table}[t]
\caption{\label{tab:eff} {\it Comparison of average SDR, number of parameters and training time per instrument.}}
\vspace{2mm}
\centering{
\begin{tabular}{ c | c  c  c} 
\hline
\multirow{2}{*}{Method}	& avg. SDR	&	\# of param. &	training time \\ & [dB] & [million] & [hour]	\\
\hline\hline
BLSTM \cite{Uhlich17} & 3.75	&	6.08	& 333 \\ 
BLEND \cite{Uhlich17} & 3.97	&	8.71	& 471 \\ 
\hline
MDenseNet		& 3.84	&	0.16	& 37 \\
MMDenseNet+ 	& 4.94	&	0.31	& 79 \\
\hline
\end{tabular}
}
\end{table} 

\section{Conclusion}
\label{sec:concl}
In this paper, we extended DenseNet to tackle the audio source separation problem. The proposed architectures have dense blocks at multiple scales connected though down-sampling and up-sampling layers, which enable the network to efficiently model both fine-grained local structure and global structure. Furthermore, we proposed a multi-band DenseNet to enable kernels in convolution layer to learn more effectively; this showed considerable performance improvement. Experimental results on the SiSEC 2016 DSD100 dataset shows that our approach outperforms the state-of-the-art by a large margin, while reducing the model size and training time significantly.

\bibliographystyle{IEEEtran}
\bibliography{mybib}

\begin{thebibliography}{10}
\providecommand{\url}[1]{#1}
\def\UrlFont{\rmfamily}
\providecommand{\newblock}{\relax}
\providecommand{\bibinfo}[2]{#2}
\providecommand\BIBentrySTDinterwordspacing{\spaceskip=0pt\relax}
\providecommand\BIBentryALTinterwordstretchfactor{4}
\providecommand\BIBentryALTinterwordspacing{\spaceskip=\fontdimen2\font plus
\BIBentryALTinterwordstretchfactor\fontdimen3\font minus
  \fontdimen4\font\relax}
\providecommand\BIBforeignlanguage[2]{{%
\expandafter\ifx\csname l@#1\endcsname\relax
\typeout{** WARNING: IEEEtran.bst: No hyphenation pattern has been}%
\typeout{** loaded for the language `#1'. Using the pattern for}%
\typeout{** the default language instead.}%
\else
\language=\csname l@#1\endcsname
\fi
#2}}

\bibitem{DuongVG10}
N.~Q.~K. Duong, E.~Vincent, and R.~Gribonval, ``Under-determined reverberant
  audio source separation using a full-rank spatial covariance model,''
  \emph{{IEEE} Trans. Audio, Speech {\&} Language Processing}, vol.~18, no.~7,
  pp. 1830--1840, 2010.

\bibitem{FitzgeraldLB16}
D.~Fitzgerald, A.~Liutkus, and R.~Badeau, ``{PROJET} - spatial audio separation
  using projections,'' in \emph{2016 {IEEE} International Conference on
  Acoustics, Speech and Signal Processing, {ICASSP} 2016, Shanghai, China,
  March 20-25, 2016}, 2016, pp. 36--40.

\bibitem{LiutkusFB15}
A.~Liutkus, D.~Fitzgerald, and R.~Badeau, ``Cauchy nonnegative matrix
  factorization,'' in \emph{{IEEE} Workshop on Applications of Signal
  Processing to Audio and Acoustics, {WASPAA} New Paltz, NY, USA}, 2015, pp.
  1--5.

\bibitem{Roux15}
J.~LeRoux, J.~R. Hershey, and F.~Weninger, ``{Deep NMF for speech
  separation},'' in \emph{Proc. ICASSP}, 2015, p. 66–70.

\bibitem{MitsufujiKS16}
Y.~Mitsufuji, S.~Koyama, and H.~Saruwatari, ``Multichannel blind source
  separation based on non-negative tensor factorization in wavenumber domain,''
  in \emph{2016 {IEEE} International Conference on Acoustics, Speech and Signal
  Processing, {ICASSP} 2016, Shanghai, China, March 20-25, 2016}, 2016, pp.
  56--60.

\bibitem{LiutkusFRPD14}
A.~Liutkus, D.~Fitzgerald, Z.~Rafii, B.~Pardo, and L.~Daudet, ``Kernel additive
  models for source separation,'' \emph{{IEEE} Trans. Signal Processing},
  vol.~62, no.~16, pp. 4298--4310, 2014.

\bibitem{OzerovF10}
A.~Ozerov and C.~F{\'{e}}votte, ``Multichannel nonnegative matrix factorization
  in convolutive mixtures for audio source separation,'' \emph{{IEEE} Trans.
  Audio, Speech {\&} Language Processing}, vol.~18, no.~3, pp. 550--563, 2010.

\bibitem{LiutkusFR15}
A.~Liutkus, D.~Fitzgerald, and Z.~Rafii, ``Scalable audio separation with light
  kernel additive modelling,'' in \emph{2015 {IEEE} International Conference on
  Acoustics, Speech and Signal Processing, {ICASSP} 2015, South Brisbane,
  Queensland, Australia, April 19-24, 2015}, 2015, pp. 76--80.

\bibitem{FitzgeraldLB162}
D.~Fitzgerald, A.~Liutkus, and R.~Badeau, ``Projection-based demixing of
  spatial audio,'' \emph{{IEEE/ACM} Trans. Audio, Speech {\&} Language
  Processing}, vol.~24, no.~9, pp. 1560--1572, 2016.

\bibitem{Nugraha15}
A.~A. Nugraha, A.~Liutkus, and E.~Vincent, ``{Multichannel music separation
  with deep neural networks},'' in \emph{Proc. EUSIPCO}, 2015.

\bibitem{Uhlich15}
S.~Uhlich, F.~Giron, and Y.~Mitsufuji, ``{Deep neural network based instrument
  extraction from music},'' in \emph{Proc. ICASSP}, 2015, pp. 2135--2139.

\bibitem{Uhlich17}
S.~Uhlich, M.~Porcu, F.~Giron, M.~Enenkl, T.~Kemp, N.~Takahashi, and
  Y.~Mitsufuji, ``{Improving Music Source Separation Based On Deep Networks
  Through Data Augmentation And Augmentation And Network Blending},'' in
  \emph{Proc. ICASSP}, 2017, pp. 261--265.

\bibitem{LeCun1998}
Y.~LeCun, L.~Bottou, Y.~Bengio, and P.~Haffner, ``{Gradient based learning
  applied to document recognition},'' in \emph{Proc. of the IEEE}, vol.~86,
  no.~11, 1998, pp. 2278--2324.

\bibitem{Sercu2015}
T.~Sercu, C.~Puhrsch, B.~Kingsbury, and Y.~LeCun, ``{Very deep multilingual
  convolutional neural networks for LVCSR},'' in \emph{Proc. ICASSP}, 2016, pp.
  4955--4959.

\bibitem{Takahashi2016}
N.~Takahashi, M.~Gygli, B.~Pfister, and L.~{Van Gool}, ``{Deep Convolutional
  Neural Networks and Data Augmentation for Acoustic Event Detection},'' in
  \emph{Proc. Interspeech}, 2016.

\bibitem{Korzeniowski16}
F.~Korzeniowski and G.~Widmer, ``{A fully convolutional deep auditory model for
  musical chord recognition},'' in \emph{Proc. International Workshop on
  Machine Learning for Signal Processing (MLSP)}, 2016.

\bibitem{Takahashi2017}
N.~Takahashi, M.~Gygli, and L.~{Van Gool}, ``Aenet: Learning deep audio
  features for video analysis,'' \emph{arXiv preprint arXiv:1701.00599}, 2017.

\bibitem{He2016}
K.~He, X.~Zhang, S.~Ren, and J.~Sun, ``{Deep residual learning for image
  recognition},'' in \emph{Proc. CVPR}, 2016.

\bibitem{Srivastava15}
R.~K. Srivastava, K.~Greff, and J.~Schmidhuber, ``{Training very deep
  networks},'' in \emph{NIPS}, 2015.

\bibitem{Huang2016}
G.~Huang, Z.~Liu, and K.~Q. Weinberger, ``Densely connected convolutional
  networks,'' \emph{arXiv preprint arXiv:1608.06993}, 2016.

\bibitem{Krizhevsky09}
A.~Krizhevsky and G.~Hinton, ``Learning multiple layers of features from tiny
  images,'' \emph{Tech Report}, 2009.

\bibitem{Netzer11}
Y.~Netzer, T.~Wang, A.~Coates, A.~Bissacco, B.~Wu, , and A.~Y. Ng, ``Reading
  digits in natural images with unsupervised feature learning,'' in \emph{NIPS
  Workshop on Deep Learning and Unsupervised Feature Learning}, 2011.

\bibitem{Liutkus17}
A.~Liutkus, F.-R. St\"{o}ter, Z.~Rai, D.~Kitamura, B.~Rivet, N.~Ito, N.~Ono, ,
  and J.~Fontecave, ``{The 2016 Signal Separation Evaluation Campaign},'' in
  \emph{Proc. LVA/ICA}, 2017, pp. 66–--70.

\bibitem{Ioffe15}
S.~Ioffe and C.~Szegedy, ``{ Batch normalization: Accelerating deep network
  training by reducing internal covariate shift},'' in \emph{Proc. ICML}, 2015.

\bibitem{Glorot11}
X.~Glorot, A.~Bordes, and Y.~Bengio, ``{Deep sparse rectifier neural
  networks},'' in \emph{Proc. AISTATS}, 2011.

\bibitem{Abdel-Hamid2013}
O.~Abdel-Hamid, L.~Deng, and D.~Yu, ``{Exploring Convolutional Neural Network
  Structures and Optimization Techniques for Speech Recognition},'' in
  \emph{Proc. Interspeech}, 2013.

\bibitem{Tieleman2012}
T.~Tieleman and G.~Hintion, ``rmsprop adaptive learning,''
  \emph{Coursera:Neural Networks for Machine Learning}, 2012.

\bibitem{Bittner14}
R.~M. Bittner, J.~Salamon, M.~Tierney, C.~C. M.~Mauch, , and J.~P. Bello,
  ``{“MedleyDB: A multitrack dataset for annotation-intensive MIR
  research},'' in \emph{Proc.ISMIR}, 2014, pp. 66–--70.

\bibitem{Vincent06}
E.~Vincent, R.~Gribonval, and C.~F\'{e}votte, ``Performance measurement in
  blind audio source separation,'' \emph{IEEE Trans. on Audio, Speech and
  Language Processing}, no.~4, pp. 1462--1469, 2006.

\end{thebibliography}
%
%
%
%
%
%
%
%
%

\end{sloppy}
\end{document}